\documentclass[twocolumn,amsmath,amssymb,floatfix,pfra,aps]{revtex4-1}

\newcommand{\ket}[1]{|#1\rangle}

\usepackage{graphicx}
\begin{document}
\title{A synthetic electric force acting on neutral atoms}
\author{Y.-J.~Lin$^1$}
\author{R.~L.~Compton$^1$}
\author{K.~Jim$\acute{\rm e}$nez-Garc$\acute{\rm i}$a$^{1,2}$}
\author{W.~D.~Phillips$^1$}
\author{J.~V.~Porto$^1$}
\author{I.~B.~Spielman$^1$}

\affiliation{$^1$Joint Quantum Institute, National Institute of
Standards and Technology, and University of Maryland, Gaithersburg,
Maryland, 20899, USA} \affiliation{$^2$Departamento de F\'{\i}sica,
Centro de Investigaci\'{o}n y Estudios Avanzados del Instituto
Polit\'{e}cnico Nacional, M\'{e}xico D.F., 07360, M\'{e}xico}

\date{\today}

\begin{abstract}
Electromagnetism is a simple example of a gauge theory where the
underlying potentials -- the vector and scalar potentials -- are
defined only up to a gauge choice. The vector potential generates
magnetic fields through its spatial variation and electric fields
through its time-dependence.  We experimentally produce a synthetic
gauge field that emerges only at low energy in a rubidium
Bose-Einstein condensate: the neutral atoms behave as charged
particles do in the presence of a homogeneous effective vector
potential. We have generated a synthetic electric field through the
time dependence of an effective vector potential, a physical
consequence even though the vector potential is spatially uniform.
\end{abstract}

\maketitle Gauge theories play a central role in modern quantum
physics. In some cases, they can be viewed as emerging as the low
energy description of a more complete theory~\cite{Levin05,Hu2009}.
Electromagnetism is the best known gauge theory and its gauge fields
are the ordinary scalar and vector potentials.  Magnetic fields
arise only from spatial variations (curl) of the vector potential,
while electric fields arise from both time variations of the vector
potential and gradients of the scalar potential; these potentials
are defined only to within a gauge choice. For a charged particle in
such potentials the canonical momentum (the variable canonically
conjugate to position) and the mechanical momentum (mass times the
velocity) are not equal. Our experiments~\cite{Lin2009a} have
realized a particular version ~\cite{Spielman2009} of a class of
proposals~\cite{Juzeliunas2006,Gunter2009,Cheneau2008,Gerbier2010,Juzeliunas2010,Ruseckas2005}
to generate effective vector potentials for neutral atoms via
interactions with laser light, and have created synthetic magnetic
fields~\cite{Lin2009b} on the success of simulating charged
condensed matter systems with neutral
atoms~\cite{Cooper2008,bloch08}. Here we demonstrate the
complementary phenomenon: the generation of a synthetic electric
field from a time-dependent effective vector potential.
Additionally, we make independent measurements of both the
mechanical momentum and canonical momentum, where the latter is
usually not possible.

\paragraph*{Introduction} The electromagnetic vector potential ${\bf A}$ for
a charged particle appears in the Hamitonian $H=({\bf p_{\rm
can}}-q{\bf A})^2/2m$, where ${\bf p_{\rm can}}$ is the canonical
momentum, $q$ is the charge, and $m$ is the mass. (The quantity
${\bf p_{\rm can}} -q{\bf A}=m{\bf v}$ is the mechanical momentum
for a particle moving with velocity $\bf{v}$.) We recently
demonstrated a technique to engineer Hamiltonians of this form for
ultra-cold atoms, and prepared a Bose-Einstein condensate (BEC) at
rest with an effective vector potential ${\bf A_0} = A_x\hat x$
constant in time and space~\cite{Lin2009a}, corresponding to ${\bf
E}={\bf B}=0$. In Ref.~\cite{Lin2009b}, we made ${\bf A}$ depend on
position giving ${\bf B}=\nabla\times{\bf A}\neq0$ but ${\bf E}=0$.
Here we add time dependence to a spatially uniform vector potential
${\bf A}(t) = A(t)\hat x$, generating a synthetic electric field
$E(t)\hat x= -\partial {\bf A}/\partial t$. The resulting force is
distinct from that due to fields arising from gradients of scalar
potentials $\phi({\bf r})$ like gravity or an external trapping
potential.  A revealing analog to which we will return is that of an
infinite solenoid of radius $r_0$ as pictured in Fig.~1A: a magnetic
field ${\bf B}=B\hat z$ exists only inside the coil, however, a
non-zero cylindrically symmetric vector potential ${\bf A}=B r_0^2
\hat \phi /2r$ in the symmetric gauge extends outside the coil. Far
from the coil, ${\bf A}$ is nearly uniform, analogous to our uniform
effective vector potential.
\begin{figure*}
\begin{center}
\includegraphics[width=4.6in]{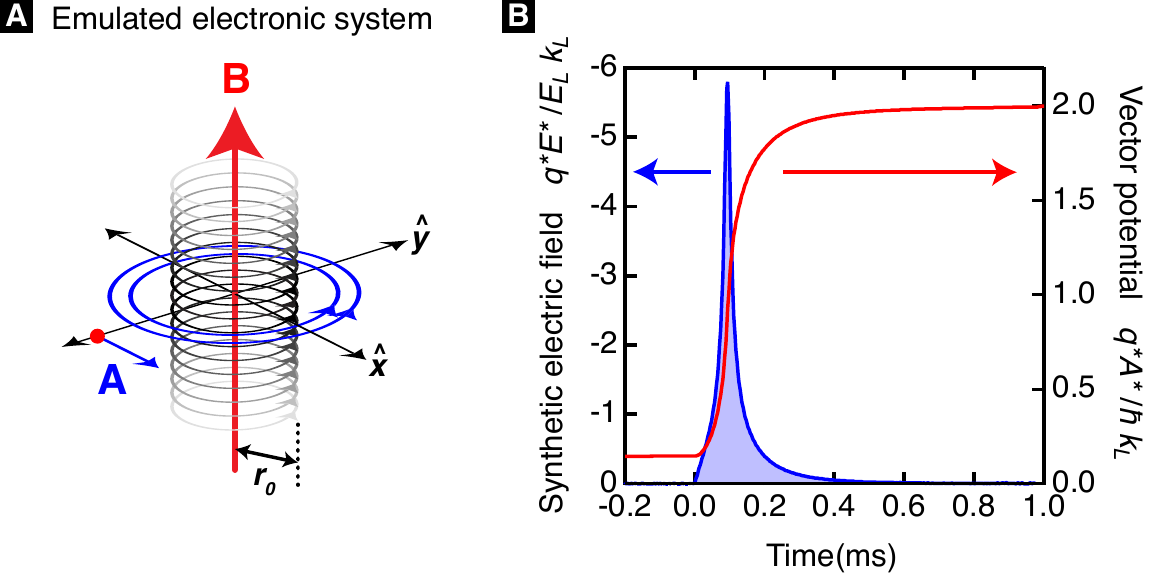}
\end{center}
\caption{Schematic of the electric field generated by a time varying
vector potential. ($\bf A$) Emulated system, showing the electric
current flowing counter clockwise in the infinite solenoid (black
coil) with radius $r_0$ and the real magnetic field ${\bf B}$ only
inside the solenoid. The blue lines represent the vector potential
${\bf A}$. A charged particle (red dot) located far from the coil
experiences a nearly uniform ${\bf A}$. ($\bf B$) Calculated time
response of the synthetic vector potential and electric field for
neutral atoms in our first measurement (see Fig.~3). The calculation
includes the known inductive response time of the bias field $B_0$,
which sets the detuning, and the calibration of detuning to vector
potential shown in Fig.~2D.} \label{Fig1}
\end{figure*}

We experimentally synthesize electromagnetic fields for neutral
atoms by illuminating an atomic $^{87}$Rb BEC with two intersecting
laser beams (Fig.~2A) that couple together three atomic spin states
within the $5S$ electronic ground state. The three new energy
eigenstates, or ``dressed states,'' are superpositions of the
uncoupled spin and linear-momentum states and have energy-momentum
dispersion relations that can be dramatically different from those
of uncoupled atoms. The atoms in a dressed state act as particles
with a single well-defined velocity ${\bf v}$, which is the
population-weighted average of all three spin components.

The dispersion relation of the lowest energy dressed state changes
near its minimum, from ${\bf p}^2/2m$ to $({\bf p}-{\bf p}_{\rm
min})^2/2m^*$ (as shown in Fig.~2C), where the minimum location
${\bf p}_{\rm min}$ plays the role of $q{\bf A}$. In addition, the
mass $m$ is modified to an effective mass  $m^*$, $m^*>m$, and both
${\bf p}_{\rm min}$ and $m^*$ are under experimental control
(although not completely independently). We identify ${\bf p}_{\rm
min}=q^*{\bf A}^*$, the product of an effective charge $q^*$ and an
effective vector potential $A^*$ for the dressed neutral atoms. As
we change ${\bf A}^*$ in time, we induce a synthetic electric field
${\bf E}^*=-\partial {\bf A}^*/\partial t$, and the dressed BEC
responds as $d(m^*{\bf v})/dt=-{\bf\nabla}\phi({\bf r})+q^*{\bf
E}^*$, where ${\bf v}$ is the velocity of the dressed atoms and
$m^*{\bf v}={\bf p_{\rm can}}-q^*{\bf A}^*$. Here, $\Delta (m^*{\bf
v})=-q^*({\bf A}^*_f-{\bf A}^*_i)$ is the momentum impulse imparted
by the synthetic electric force $q^*{\bf E}^*$.

We study the physical consequences of sudden temporal changes of the
effective vector potential for the dressed BEC. These changes are
always adiabatic such that the BEC remains in the same dressed
state. We measure the resulting change of the BEC's momentum, which
is in complete quantitative agreement with our calculations and
constitutes the first observation of synthetic electric fields for
neutral atoms.
\begin{figure*}
\begin{center}
\includegraphics[width=4.6in]{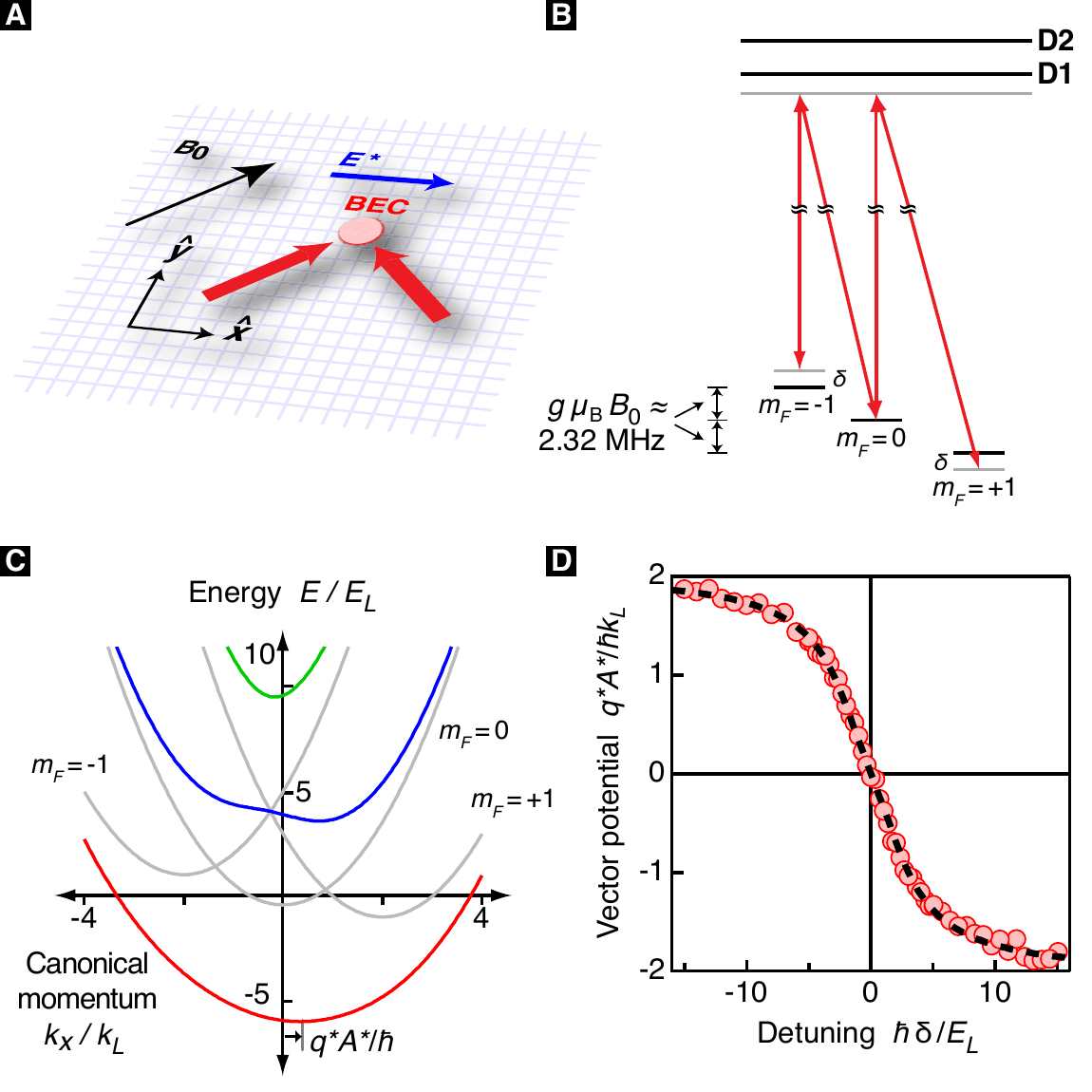}
\end{center}
\caption{Experiment setup for synthetic electric fields. ($\bf A$)
Physical implementation indicating the two Raman laser beams
incident on the BEC (red arrows) and the physical bias magnetic
field $B_0$ (black arrow). The blue arrow indicates the direction of
the synthetic electric field $E^*$. ($\bf B$) The three $m_F$ levels
of the $F=1$ ground state manifold are shown as coupled by the Raman
beams. (C) The dressed state eigenenergies as a function of
canonical momentum for the realized coupling strength of $\hbar
\Omega_R=10.5 E_L$ at a representative detuning $\hbar \delta=
-1E_L$ (colored curves). The grey curves show the energies of the
uncoupled states, and the red curve depicts the lowest energy
dressed state in which we load the BEC. The black arrow indicates
the dressed BEC's canonical momentum $p_{\rm can}=q^*A^*$, where
$A^*$ is the vector potential. (D) Measured vector potential from
the canonical momentum.} \label{Fig2}
\end{figure*}
\paragraph*{Experimental realization} Our system, depicted in
Fig.~2A, consists of a $F=1$ $^{87}{\rm Rb}$ BEC with about
$1.4\times10^5$ atoms initially at rest~\cite{Lin2009,hung08}; a
small physical magnetic field $B_0$ Zeeman-shifts each of the three
magnetic sublevels $m_F=0,\pm1$ by an amount $E_{0,\pm1}$, but
produces no Lorentz force on our charge-neutral BEC. Our experiment
is performed in the linear Zeeman regime, i.e.,
$E_{-1}\approx-E_{+1}\approx g\mu_B B_0\gg |E_{0}|$, with linear and
quadratic Zeeman shifts $\hbar\omega_{\rm
Z}=(E_{-1}-E_{+1})/2\approx h\times2.32\ {\rm MHz}$ and
$-\hbar\epsilon=E_0-(E_{-1}+E_{+1})/2\approx -h\times784\ {\rm Hz}$.
In addition, a pair of $\lambda=801\ {\rm nm}$ laser beams,
intersecting at $90^\circ$ at the BEC, couples the $m_F$ states with
strength $\Omega_R$. These Raman lasers differ in frequency by
$\Delta\omega_L \approx \omega_{\rm Z}$ and we define the detuning
from Raman resonance as $\delta=\Delta\omega_L-\omega_{\rm Z}$. In
our experiments $\hbar\Omega_R\approx 10E_L$ and $|\hbar\delta| < 60
E_L$, where $E_L=\hbar^2 k_L^2/2 m$ and $k_L = \sqrt{2}\pi/\lambda$
are the appropriate units of energy and momentum.

When the atoms are rapidly moving or the Raman lasers are far from
resonance ($k_L v$ or $\delta \gg \Omega_R$), the coupling lasers
hardly affect the atoms. However, nearly resonant, slowly moving
atoms are strongly dressed: the three uncoupled states transform
into three new dressed states. The spin and linear-momentum state
$\ket{k_x,m_F=0}$ is coupled to states $\ket{k_x-2k_L,m_F=+1}$ and
$\ket{k_x+2k_L,m_F=-1}$, where $\hbar k_x$ is the momentum of
$\ket{m_F=0}$ along $\hat{x}$, and $2\hbar k_L\hat x$ is the
momentum difference between two Raman beams. In the frame rotating
at $\Delta \omega_L$ under the rotating wave approximation, the
Hamiltonian matrix $H(k_x)/\hbar$ for motion along $\hat x$ is

$$\left(
\begin{array}{ccc} \frac{\hbar}{2m}(k_x+2k_L)^2-\delta & \Omega_R/2 & 0 \\
\Omega_R/2 & \frac{\hbar}{2m} k_x^2-\epsilon & \Omega_R/2 \\
0 & \Omega_R/2 & \frac{\hbar}{2m}(k_x-2k_L)^2+\delta\end{array}
\right).
$$
Diagonalization for each $k_x$ yields three dressed eigenstates,
each of which is a superposition of the uncoupled
states~\cite{Lin2009a}, with energies $E_j(k_x)$ shown in Fig.~2C
(gray for uncoupled states, colored for dressed states); we focus on
atoms in the lowest energy energy-momentum (red) curve. When the
atoms' energy (interaction and kinetic) is small compared to the
$\approx10 E_L$ energy difference between the curves--as it is in
our experiments--the contact-interacting atoms will remain within
the lowest energy dressed state~\cite{Spielman2009}. Such low energy
particles do not reveal their spin and momentum components.
\begin{figure}
\begin{center}
\includegraphics[width=2.3in]{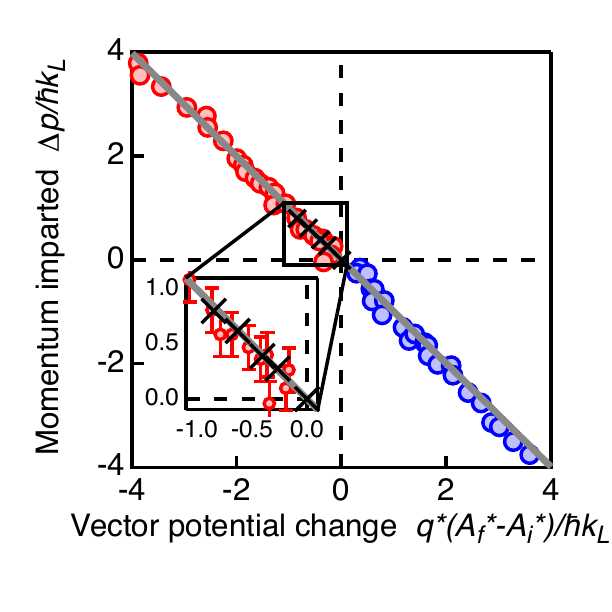}
\end{center}
\caption{Change in momentum from the synthetic electric field kick.
This figure displays three distinct sets of data shown by red and
blue circles and black crosses.  The data were obtained by applying
a synthetic electric field by changing the vector potential from
$q^*A^*_i$ (between $+2\hbar k_{\rm L}$ and $-2\hbar k_{\rm L}$) to
$q^*A^*_f$.  Circles indicate data where the external trap was
removed right before the change in $A^*$ : for red (blue) symbols
$q^*A^*_f=-2\hbar k_{\rm L} (+2\hbar k_{\rm L})$. The black crosses,
more visible in the inset, show the amplitude of canonical momentum
oscillations when the trapping potential was left on after the
field-kick.  The typical uncertainties are also visible in the
inset. The grey line is a linear fit to the data (circles) yielding
slope $-0.996\pm0.008$, where the expected slope is $-1$.}
\label{Fig3}
\end{figure}
In the low energy limit, dressed atoms have a new effective
Hamiltonian for motion along $\hat x$, $H_x=(\hbar
k_x-q^*A^*_x)^2/2m^*$ (motion along $\hat y$ and $\hat z$ is
unaffected), where we have chosen a gauge such that the momentum of
the $m_F=0$ component $\hbar k_x \equiv p_{\rm can}$ is the
canonical momentum of the dressed state. Our approach introduces the
term $q^*A^*_x$, which depends on the detuning $\delta$ as shown in
Fig.~2D, and an effective mass $m^*$. The red curve in Fig.~2C shows
$H_x$ for $q^*A^*_x>0$ and the arrow indicates the canonical
momentum of a BEC in the lowest energy dressed state, $p_{\rm
can}=q^*A^*_x$. Although this dressed BEC is at rest ($v=\partial
H_x/\partial \hbar k_x=0$, zero group velocity), it is composed of
three bare spin states each with a different momentum, among which
the momentum of $|m_F=0\rangle$ is $\hbar k_x=p_{\rm can}$. Although
none of its three bare spin components has zero momentum, the BEC's
averaged momentum is zero.

We measure $p_{\rm can}$ by first removing the coupling fields and
trapping potentials and then allowing the atoms to freely expand for
a $t=20.1\ {\rm ms}$ time-of-flight (TOF). Because the three
components of the dressed state
$\{\ket{k_x,m_F=0},\ket{k_x\mp2k_L,m_F=\pm1}\}$ differ in momentum
by $\pm \hbar 2k_L$, they quickly separate.

Using a combination of adiabatically applied Raman and rf
(radio-frequency) fields we transfer (load) the BEC initially in
$\ket{m_F=-1}$ into the lowest energy dressed state with a ${\bf
A^*}=A^* \hat x$ (see Ref.~\cite{Lin2009a} for a complete technical
discussion of loading), i.e., with a canonical momentum $p_{\rm
can}=q^*A^*$. We measure $q^* A^*$ as the momentum of
$|m_F=0\rangle$; Fig.~2D shows how the measured and predicted $A^*$
depend on the detuning $\delta$. With this calibration, we use
$\delta$ to control the effective vector potential.

\paragraph*{Synthetic electric field} Electric fields result either
from spatially varying scalar potentials or from time varying vector
potentials. An example of the latter case is illustrated in Fig.~1A
for an infinite solenoid. The magnetic field ${\bf B}$ exits only
inside the coil; when the current is constant the electric field
${\bf E}$ is everywhere zero but the symmetric gauge vector
potential is zero only at the solenoid center. When the current is
changed in a time interval $\Delta t$ the magnetic field inside the
coil changes to a final value $B(\Delta t)$
, and the exterior vector potential changes by $\Delta {\bf
A}=\left[B(\Delta t)-B(0)\right]r_0^2\hat \phi/2r$.  Here, a charged
particle on the $\hat y$ axis feels an electric field $-(\partial
A/\partial t)\hat x$ during the time $\Delta t$, leading to a
mechanical momentum kick $\Delta {\bf p} = -q \Delta A \hat x$, even
outside the solenoid.

We realize a synthetic electric field $E^*$ for ultracold neutral
atoms by changing the effective vector potential from an initial
value $A^*_i$ to a final value $A^*_f$. We perform two types of
measurements of $E^*$: (1) we remove the trapping potential right
before we change $A^*$ such that $E^*$ can accelerate the atoms
unimpeded, and measure the resulting change from zero of the BEC's
velocity; and (2) after changing $A^*$, we allow the trapped,
dressed atoms to oscillate in their harmonic potential, and measure
the magnitude and frequency of the oscillations of both the
canonical momentum and velocity.

We prepare our BEC at rest with an initial vector potential $A^*_i
{\hat x}$. In the first measurement, we change the detuning $\delta$
in $0.8\ {\rm ms}$, leading to a change of the vector potential to
$A^*_f$, after which the Raman coupling is turned off in 0.2 ms. As
$A^*$ changes from $A^*_i$ to $A^*_f$, a synthetic electric field
$E^*$ appears and accelerates the BEC. We image the atoms after a
TOF and record their final velocities. Figure~3 shows the momentum
$\Delta p$ imparted to the atoms by $E^*$ as a function of the
vector potential change $q^*(A^*_f-A^*_i)$, denoted by red (blue)
symbols with $q^*A^*_f/\hbar=2k_{\rm L} (-2k_{\rm L})$. We have
chosen a sufficiently large final detuning $\hbar \delta= \mp 60 E_L
$ such that the final atomic state is a nearly pure spin state,
$|m_F=+1\rangle$ for $q^*A^*_f/\hbar=2k_{\rm L}$ or $|m_F=-1\rangle$
for $q^*A^*_f/\hbar=-2k_{\rm L}$. For this essentially undressed
final state, $m^*=m$ and the imparted momentum is $\Delta
p=m^*v=mv$, equal to the change in mechanical momentum. We performed
a linear fit $\Delta p = Cq^* (A^*_f-A^*_i)$ to the data and obtain
$C=-0.996(8)$, in good agreement with the expected $\Delta p= -q^*
(A^*_f-A^*_i)$.

\begin{figure*}
\begin{center}
\includegraphics[width=4.6in]{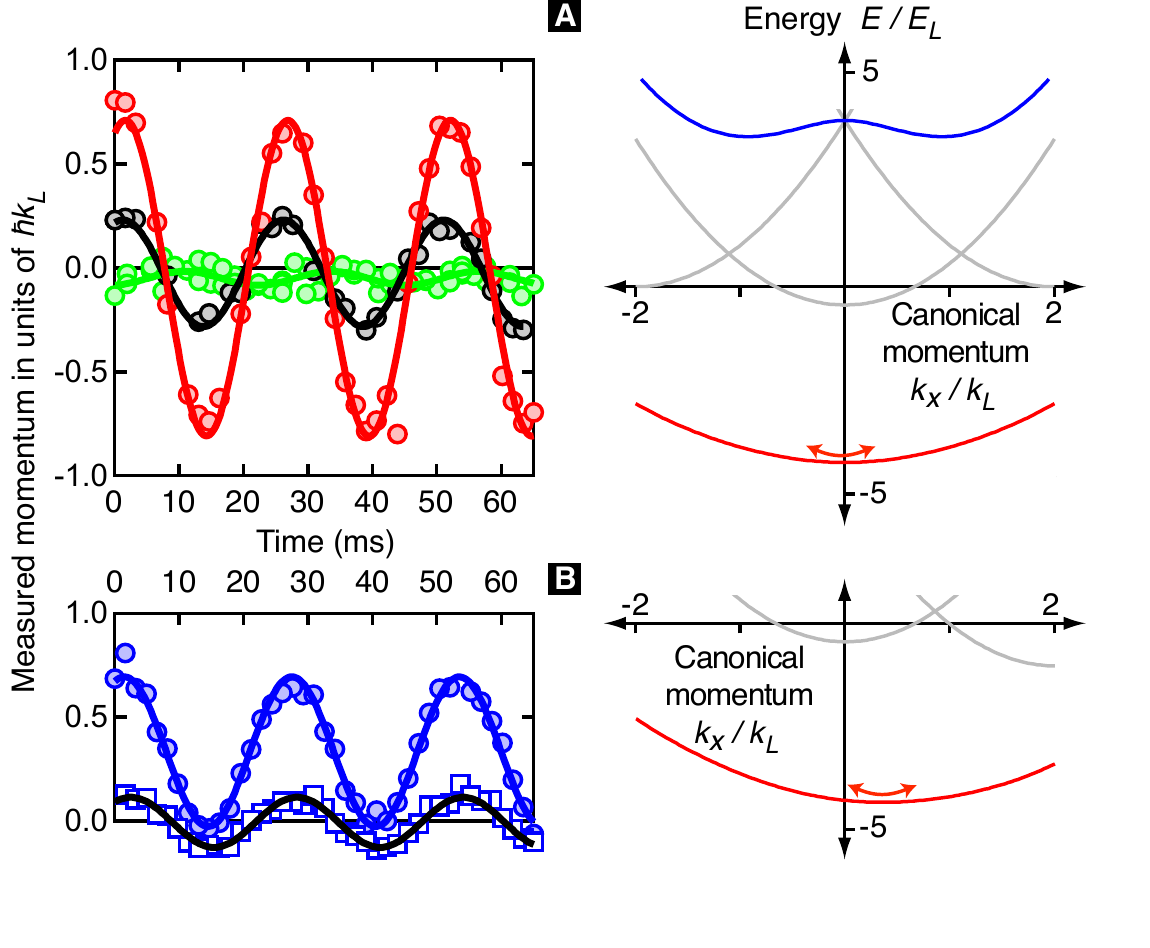}
\end{center}
\caption{Oscillating atoms in the trapping potential after
application of a synthetic electric field pulse. $(\bf A)(\bf B)$
Left panels: The vector potential is changed from
$q^*A^*_i=0.75~\hbar k_{\rm L}$ (red circles), $0.25~\hbar k_{\rm
L}$ (black circles), and $q^*A^*_i=q^*A^*_f \approx 0$ (green
circles), respectively, all to $q^*A^*_f$, and from $q^*A^*_i=0.75~
\hbar k_{\rm L}$ (blue circles) to $q^*A^*_f=0.35~ \hbar k_{\rm L}$
(blue squares). The measured momentum for all circles is the
canonical momentum $p_{\rm can}$, and that for the square is the
mechanical momentum $mv$. In (B), $p_{\rm can}$ oscillates about
$q^*A^*_f \neq 0$ while $mv$ oscillates about zero. Right panels:
Energy-momentum dispersion curves for uncoupled states (grey) and
dressed states (colored). The arrows indicate oscillations of
$p_{\rm can}$ about $q^*A^*_f$ for atoms in the lowest energy
dressed state.} \label{Fig4}
\end{figure*}
In the measurement described above, the final value of $A^*=A^*_f$
resulted from a detuning so large that the final-state atoms were
essentially undressed. In a second measurement, we examined the time
evolution of atoms that remain trapped and strongly dressed. Here we
again changed $A^*$ from $A^*_i$ to $A^*_f$ in about $\Delta t=0.3\
{\rm ms}$ but left the dressed BEC in the harmonic confining
potential for a variable time before abruptly removing the confining
potential and the Raman fields for a TOF. As the BEC oscillated in
the trap, we monitored the out-of-equilibrium canonical momentum
$p_{\rm can}$. It is our access to the internal degrees of freedom
-- here projectively measuring the composition of the Raman dressed
state -- that allows the determination of $p_{\rm can}$. Figure 4A
shows the resulting time evolution of $p_{\rm can}$ for several
different initial $A^*_i$ all for $A^*_f \approx 0$; as expected, we
observe oscillations of $p_{\rm can}$ about $q^*A^*_f$. Since
$\Delta t$ is small compared to our $\approx 25$~ms trap period, we
expect the change of momentum to be dominated by $\Delta
p=-q^*(A^*_f-A^*_i)$, where the contribution from the trapping force
as $A^*$ changes is negligible. This translates into an oscillation
amplitude $\Delta p$ in both $p_{\rm can}$ and $m^*v=p_{\rm
can}-q^*A^*_f$ of dressed atoms; the solid crosses in Fig.~3 show
the amplitude of the sinusoidal oscillations in $p_{\rm can}$ as a
function of $A^*_f-A^*_i\approx -A^*_i$, proving that the synthetic
electric field has imparted the expected momentum kick.

We repeated the experiment with a nonzero $q^*A^*_f/\hbar\approx
0.35 k_{\rm L}$, and observed, remarkably, the sinusoidal
oscillations in $p_{\rm can}$ were offset from zero as depicted in
Fig.~4B. $p_{\rm can}$ evolves smoothly about $q^*A^*_f\neq0$,
illustrating the observed quantity is not the mechanical momentum
$mv$, which should oscillate about zero. We also measured $mv$,
where $v$ is the population-weighted average velocity of all spin
components; while $mv$ indeed oscillates about zero, the oscillation
amplitude is smaller than that of $p_{\rm can}$. Given an increased
effective mass $m/m^* \approx 0.4$, the trap frequency $\nu_x$ along
$\hat x$ was reduced by $\sqrt{m/m^*}$ from that for undressed
atoms, and the oscillation amplitude of $mv$ should be reduced by
$m/m^*$ from that of $p_{\rm can}$. Our results show a reduction
factor of $0.38(4)$ in $\nu_x^2$, in agreement with the predicted
$m/m^*$, and the reduction factor of $0.30(2)$ in the momentum
oscillation amplitude is slightly smaller than predicted.

\paragraph*{Conclusion}
We have demonstrated how a time-dependent, spatially uniform
effective vector potential for low-energy dressed neutral atoms
gives rise to a synthetic electric force. We have measured the
effect of such forces for both trapped and untrapped atoms, and for
the final state both dressed and undressed. While the electric
forces generated here are spatially homogenous, this technique is
generally applicable to create spatially varying forces. Indeed,
since the effective vector potential $A^*$ is parameterized by the
Raman detuning $\delta$, it can be locally patterned via suitable
spatially inhomogeneous magnetic fields or vector light shifts,
giving rise to a nearly arbitrary range of time-dependent
potentials. For azimuthal vector potentials, our method of measuring
both the mechanical and canonical momentum from the induced electric
fields can be used to identify the superfluid fraction of a
BEC~\cite{cooper10}, or any other cold-atom systems.

This work was partially supported by ONR, ARO with funds from the
DARPA OLE program, and the NSF through the JQI Physics Frontier
Center. R.L.C. acknowledges the NIST/NRC postdoctoral program and
K.J.-G thanks CONACYT.


\begin{thebibliography}{10}%
\makeatletter
\providecommand \@ifxundefined [1]{%
 \ifx #1\undefined \expandafter \@firstoftwo
 \else \expandafter \@secondoftwo
\fi
}%
\providecommand \@ifnum [1]{%
 \ifnum #1\expandafter \@firstoftwo
 \else \expandafter \@secondoftwo
\fi
}%
\providecommand \enquote [1]{``#1''}%
\providecommand \bibnamefont  [1]{#1}%
\providecommand \bibfnamefont [1]{#1}%
\providecommand \citenamefont [1]{#1}%
\providecommand\href[0]{\@sanitize\@href}%
\providecommand\@href[1]{\endgroup\@@startlink{#1}\endgroup\@@href}%
\providecommand\@@href[1]{#1\@@endlink}%
\providecommand \@sanitize [0]{\begingroup\catcode`\&12\catcode`\#12\relax}%
\@ifxundefined \pdfoutput {\@firstoftwo}{%
 \@ifnum{\z@=\pdfoutput}{\@firstoftwo}{\@secondoftwo}%
}{%
 \providecommand\@@startlink[1]{\leavevmode}%
 \providecommand\@@endlink[0]{}%
}{%
 \providecommand\@@startlink[1]{%
  \leavevmode
  \pdfstartlink
   attr{/Border[0 0 1 ]/H/I/C[0 1 1]}%
   user{/Subtype/Link/A<</Type/Action/S/URI/URI(#1)>>}%
  \relax
 }%
 \providecommand\@@endlink[0]{\pdfendlink}%
}%
\providecommand \url  [0]{\begingroup\@sanitize \@url }%
\providecommand \@url [1]{\endgroup\@href {#1}{\urlprefix}}%
\providecommand \urlprefix [0]{URL }%
\providecommand \Eprint[0]{\href }%
\@ifxundefined \urlstyle {%
  \providecommand \doi [1]{doi:\discretionary{}{}{}#1}%
}{%
  \providecommand \doi [0]{doi:\discretionary{}{}{}\begingroup
  \urlstyle{rm}\Url }%
}%
\providecommand \doibase [0]{http://dx.doi.org/}%
\providecommand \Doi[1]{\href{\doibase#1}}%
\providecommand \bibAnnote [3]{%
  \BibitemShut{#1}%
  \begin{quotation}\noindent
    \textsc{Key:}\ #2\\\textsc{Annotation:}\ #3%
  \end{quotation}%
}%
\providecommand \bibAnnoteFile [2]{%
  \IfFileExists{#2}{\bibAnnote {#1} {#2} {\input{#2}}}{}%
}%
\providecommand \typeout [0]{\immediate \write \m@ne }%
\providecommand \selectlanguage [0]{\@gobble}%
\providecommand \bibinfo [0]{\@secondoftwo}%
\providecommand \bibfield [0]{\@secondoftwo}%
\providecommand \translation [1]{[#1]}%
\providecommand \BibitemOpen[0]{}%
\providecommand \bibitemStop [0]{}%
\providecommand \bibitemNoStop [0]{.\EOS\space}%
\providecommand \EOS [0]{\spacefactor3000\relax}%
\providecommand \BibitemShut [1]{\csname bibitem#1\endcsname}%
\bibitem{Levin05}%
  \BibitemOpen
  \bibfield{author}{%
  \bibinfo {author} {\bibfnamefont{M.}~\bibnamefont{Levin}}\ and\ \bibinfo
  {author} {\bibfnamefont{X.-G.}\ \bibnamefont{Wen}},\ }%
  \bibfield{journal}{%
  \Doi{10.1103/RevModPhys.77.871}{\bibinfo {journal} {Rev. Mod. Phys.}}\ }%
  \textbf{\bibinfo {volume} {77}},\ \bibinfo {pages} {871} (\bibinfo {year} {2005})%
  \bibAnnoteFile{NoStop}{Levin05}%
\bibitem{Hu2009}%
  \BibitemOpen
  \bibfield{author}{%
  \bibinfo {author} {\bibfnamefont{B.~L.}\ \bibnamefont{Hu}},\ }%
  \bibfield{journal}{%
  \bibinfo {journal} {J. Phys. Conf. Ser.}\ }%
  \textbf{\bibinfo {volume} {174}},\ \bibinfo {pages} {012015} (\bibinfo {year}
  {2009})%
  \bibAnnoteFile{NoStop}{Hu2009}%
\bibitem{Lin2009a}%
  \BibitemOpen
  \bibfield{author}{%
  \bibinfo {author} {\bibfnamefont{Y.-J.}\ \bibnamefont{Lin}}, \bibinfo
  {author} {\bibfnamefont{R.~L.}\ \bibnamefont{Compton}}, \bibinfo {author}
  {\bibfnamefont{A.~R.}\ \bibnamefont{Perry}}, \bibinfo {author}
  {\bibfnamefont{W.~D.}\ \bibnamefont{Phillips}}, \bibinfo {author}
  {\bibfnamefont{J.~V.}\ \bibnamefont{Porto}},\ and\ \bibinfo {author}
  {\bibfnamefont{I.~B.}\ \bibnamefont{Spielman}},\ }%
  \bibfield{journal}{%
  \Doi{10.1103/PhysRevLett.102.130401}{\bibinfo {journal} {Phys. Rev. Lett.}}\
  }%
  \textbf{\bibinfo {volume} {102}},\ \bibinfo {eid} {130401} (\bibinfo {year}
  {2009})%
  \bibAnnoteFile{NoStop}{Lin2009a}%
\bibitem{Spielman2009}%
  \BibitemOpen
  \bibfield{author}{%
  \bibinfo {author} {\bibfnamefont{I.~B.}\ \bibnamefont{Spielman}},\ }%
  \bibfield{journal}{%
  \Doi{10.1103/PhysRevA.79.063613}{\bibinfo {journal} {Phys. Rev. A}}\ }%
  \textbf{\bibinfo {volume} {79}},\ \bibinfo {eid} {063613} (\bibinfo {year}
  {2009})%
  \bibAnnoteFile{NoStop}{Spielman2009}%
\bibitem{Juzeliunas2006}%
  \BibitemOpen
  \bibfield{author}{%
  \bibinfo {author} {\bibfnamefont{G.}~\bibnamefont{Juzeli{\=u}nas}}, \bibinfo
  {author} {\bibfnamefont{J.}~\bibnamefont{Ruseckas}}, \bibinfo {author}
  {\bibfnamefont{P.}~\bibnamefont{{\"O}hberg}},\ and\ \bibinfo {author}
  {\bibfnamefont{M.}~\bibnamefont{Fleischhauer}},\ }%
  \bibfield{journal}{%
  \bibinfo {journal} {Phys. Rev. A}\ }%
  \textbf{\bibinfo {volume} {73}},\ \bibinfo {pages} {025602} (\bibinfo {year}
  {2006})%
  \bibAnnoteFile{NoStop}{Juzeliunas2006}%
\bibitem{Gunter2009}%
  \BibitemOpen
  \bibfield{author}{%
  \bibinfo {author} {\bibfnamefont{K.~J.}\ \bibnamefont{G\"{u}nter}}, \bibinfo
  {author} {\bibfnamefont{M.}~\bibnamefont{Cheneau}}, \bibinfo {author}
  {\bibfnamefont{T.}~\bibnamefont{Yefsah}}, \bibinfo {author}
  {\bibfnamefont{S.~P.}\ \bibnamefont{Rath}},\ and\ \bibinfo {author}
  {\bibfnamefont{J.}~\bibnamefont{Dalibard}},\ }%
  \bibfield{journal}{%
  \bibinfo {journal} {Phys. Rev. A}\ }%
  \textbf{\bibinfo {volume} {79}},\ \bibinfo {pages} {011604} (\bibinfo {year}
  {2009})%
  \bibAnnoteFile{NoStop}{Gunter2009}%
\bibitem{Cheneau2008}%
  \BibitemOpen
  \bibfield{author}{%
  \bibinfo {author} {\bibfnamefont{M.}~\bibnamefont{Cheneau}}, \bibinfo
  {author} {\bibfnamefont{S.~P.}\ \bibnamefont{Rath}}, \bibinfo {author}
  {\bibfnamefont{T.}~\bibnamefont{Yefsah}}, \bibinfo {author}
  {\bibfnamefont{K.~J.}\ \bibnamefont{Gunter}}, \bibinfo {author}
  {\bibfnamefont{G.}~\bibnamefont{Juzeliunas}},\ and\ \bibinfo {author}
  {\bibfnamefont{J.}~\bibnamefont{Dalibard}},\ }%
  \bibfield{journal}{%
  \bibinfo {journal} {Europhys. Lett.}\ }%
  \textbf{\bibinfo {volume} {83}},\ \bibinfo {pages} {60001 (6pp)} (\bibinfo
  {year} {2008})%
  \bibAnnoteFile{NoStop}{Cheneau2008}%
\bibitem{Gerbier2010}%
  \BibitemOpen
  \bibfield{author}{%
  \bibinfo {author} {\bibfnamefont{F.}~\bibnamefont{Gerbier}}\ and\ \bibinfo
  {author} {\bibfnamefont{J.}~\bibnamefont{Dalibard}},\ }%
  \bibfield{journal}{%
  \bibinfo {journal} {New Journal of Physics}\ }%
  \textbf{\bibinfo {volume} {12}},\ \bibinfo {pages} {033007} (\bibinfo {year}
  {2010})%
  \bibAnnoteFile{NoStop}{Gerbier2010}%
\bibitem{Juzeliunas2010}%
  \BibitemOpen
  \bibfield{author}{%
  \bibinfo {author} {\bibfnamefont{G.}~\bibnamefont{Juzeliunas}}, \bibinfo
  {author} {\bibfnamefont{J.}~\bibnamefont{Ruseckas}},\ and\ \bibinfo {author}
  {\bibfnamefont{J.}~\bibnamefont{Dalibard}},\ }%
  \bibfield{journal}{%
  \bibinfo {journal} {Phys. Rev. A}\ }%
  \textbf{\bibinfo {volume} {81}},\ \bibinfo {pages} {053403} (\bibinfo {year}
  {2010})%
  \bibAnnoteFile{NoStop}{Juzeliunas2010}%
\bibitem{Ruseckas2005}%
  \BibitemOpen
  \bibfield{author}{%
  \bibinfo {author} {\bibfnamefont{J.}~\bibnamefont{Ruseckas}}, \bibinfo
  {author} {\bibfnamefont{G.}~\bibnamefont{Juzeliunas}}, \bibinfo {author}
  {\bibfnamefont{P.}~\bibnamefont{Ohberg}},\ and\ \bibinfo {author}
  {\bibfnamefont{M.}~\bibnamefont{Fleischhauer}},\ }%
  \bibfield{journal}{%
  \bibinfo {journal} {Phys. Rev. Lett.}\ }%
  \textbf{\bibinfo {volume} {95}} (\bibinfo {year} {2005})%
  \bibAnnoteFile{NoStop}{Ruseckas2005}%
\bibitem{Lin2009b}%
  \BibitemOpen
  \bibfield{author}{%
  \bibinfo {author} {\bibfnamefont{Y.-J.}\ \bibnamefont{Lin}}, \bibinfo
  {author} {\bibfnamefont{R.~L.}\ \bibnamefont{Compton}}, \bibinfo {author}
  {\bibfnamefont{K.}~\bibnamefont{Jimenez-Garcia}}, \bibinfo {author}
  {\bibfnamefont{J.~V.}\ \bibnamefont{Porto}},\ and\ \bibinfo {author}
  {\bibfnamefont{I.~B.}\ \bibnamefont{Spielman}},\ }%
  \bibfield{journal}{%
  \bibinfo {journal} {Nature}\ }%
  \textbf{\bibinfo {volume} {462}},\ \bibinfo {pages} {628} (\bibinfo {year}
  {2009})%
  \bibAnnoteFile{NoStop}{Lin2009b}%
\bibitem{Cooper2008}%
  \BibitemOpen
  \bibfield{author}{%
  \bibinfo {author} {\bibfnamefont{N.~R.}\ \bibnamefont{Cooper}},\ }%
  \bibfield{journal}{%
  \bibinfo {journal} {Adv. Phys.}\ }%
  \textbf{\bibinfo {volume} {57}},\ \bibinfo {pages} {539 } (\bibinfo {year}
  {2008})%
  \bibAnnoteFile{NoStop}{Cooper2008}%
\bibitem{bloch08}%
  \BibitemOpen
  \bibfield{author}{%
  \bibinfo {author} {\bibfnamefont{I.}~\bibnamefont{Bloch}}, \bibinfo {author}
  {\bibfnamefont{J.}~\bibnamefont{Dalibard}},\ and\ \bibinfo {author}
  {\bibfnamefont{W.}~\bibnamefont{Zwerger}},\ }%
  \bibfield{journal}{%
  \Doi{10.1103/RevModPhys.80.885}{\bibinfo {journal} {Rev. Mod. Phys.}}\ }%
  \textbf{\bibinfo {volume} {80}},\ \bibinfo {pages} {885} (\bibinfo {year} {2008})%
  \bibAnnoteFile{NoStop}{bloch08}%
\bibitem{Lin2009}%
  \BibitemOpen
  \bibfield{author}{%
  \bibinfo {author} {\bibfnamefont{Y.-J.}\ \bibnamefont{Lin}}, \bibinfo
  {author} {\bibfnamefont{A.~R.}\ \bibnamefont{Perry}}, \bibinfo {author}
  {\bibfnamefont{R.~L.}\ \bibnamefont{Compton}}, \bibinfo {author}
  {\bibfnamefont{I.~B.}\ \bibnamefont{Spielman}},\ and\ \bibinfo {author}
  {\bibfnamefont{J.~V.}\ \bibnamefont{Porto}},\ }%
  \bibfield{journal}{%
  \Doi{10.1103/PhysRevA.79.063631}{\bibinfo {journal} {Phys. Rev. A}}\ }%
  \textbf{\bibinfo {volume} {79}},\ \bibinfo {eid} {063631} (\bibinfo {year}
  {2009})%
  \bibAnnoteFile{NoStop}{Lin2009}%
\bibitem{hung08}%
  \BibitemOpen
  \bibfield{author}{%
  \bibinfo {author} {\bibfnamefont{C.-L.}\ \bibnamefont{Hung}}, \bibinfo
  {author} {\bibfnamefont{X.}~\bibnamefont{Zhang}}, \bibinfo {author}
  {\bibfnamefont{N.}~\bibnamefont{Gemelke}},\ and\ \bibinfo {author}
  {\bibfnamefont{C.}~\bibnamefont{Chin}},\ }%
  \bibfield{journal}{%
  \bibinfo {journal} {Phys. Rev. A}\ }%
  \textbf{\bibinfo {volume} {78}},\ \bibinfo {pages} {011604} (\bibinfo {year}
  {2008})%
  \bibAnnoteFile{NoStop}{hung08}%
\bibitem{cooper10}%
  \BibitemOpen
  \bibfield{author}{%
  \bibinfo {author} {\bibfnamefont{N.~R.}\ \bibnamefont{Cooper}}\ and\ \bibinfo
  {author} {\bibfnamefont{Z.}~\bibnamefont{Hadzibabic}},\ }%
  \bibfield{journal}{%
  \bibinfo {journal} {Phys. Rev. Lett.}\ }%
  \textbf{\bibinfo {volume} {104}},\ \bibinfo {pages} {030401} (\bibinfo {year}
  {2010})%
  \bibAnnoteFile{NoStop}{cooper10}%
\bibitem{Castin96}%
  \BibitemOpen
  \bibfield{author}{%
  \bibinfo {author} {\bibfnamefont{Y.}~\bibnamefont{Castin}}\ and\ \bibinfo
  {author} {\bibfnamefont{R.}~\bibnamefont{Dum}},\ }%
  \bibfield{journal}{%
  \bibinfo {journal} {Phys. Rev. Lett.}\ }%
  \textbf{\bibinfo {volume} {77}},\ \bibinfo {pages} {5315} (\bibinfo {year}
  {1996})%
  \bibAnnoteFile{NoStop}{Castin96}%
\end{thebibliography}
%

\section*{Methods}

\subsection*{1~~Dynamic change of effective vector potentials}

In our first measurement of synthetic electric fields, we observed
the momentum imparted by the field kick, resulting from a change in
the effective vector potential from an arbitrarily chosen $q^*A^*_i$
to $q^*A^*_f=\pm2 \hbar k_{\rm L}$ (see Fig.~3). In principle we
could use any $A^*_f$ and observe a momentum kick
$q^*(A^*_i-A^*_f)$, however, in general the effective vector
potential also depends on the strength of the Raman coupling
$\Omega_R$. As a result, additional synthetic electric fields
typically appear when $\Omega_R$ is adiabatically turned off. There
are three specific cases for which $A^*$ does not depend on
$\Omega_R$: when the detuning $\delta=0$ and $\delta\rightarrow
\pm\infty$. For the former case, only when $\left|q^*A^*_i\right| <
\hbar k_L$ is there no additional electric force during the removal
of $\Omega_R$, where the final atomic state is $|m_F=0\rangle$. For
$\left|q^*A^*_i\right| > \hbar k_L$, the final atomic state is
$|m_F=\pm 1 \rangle$ with an additional momentum of $\mp 2\hbar k_L$
imparted. Thus in our experiment we changed the vector potential
from $q^*A^*_i$ to $q^*A^*_f=\pm2 \hbar k_{\rm L}$ by changing the
detuning from $\delta_i$ to a large $\hbar \delta_f=\mp60E_{L}$; the
subsequent turnoff of $\Omega_R$ then exerted no additional forces.

\subsection*{2~~Control of Raman detuning }

In all of our experiments, we set the Raman detuning $\delta$ with
small changes of the bias magnetic field away from resonance and
hold the $2.32\ {\rm MHz}$ frequency difference between the Raman
beams constant. Because all temporal changes in $\delta$ lead to
synthetic electric fields, bias magnetic field noise and relative
laser frequency noise can lead to motion in the trap or heating. We
phase locked the two Raman beams and observed no change in the
heating, showing that relative laser frequency noise is not
important in our experiment. However, our experiment is very
sensitive to ambient magnetic field noise, here tied to the 60 Hz
line. This noise gives rise to intractable dynamics of the canonical
momentum of the dressed state, where $\delta$ is held constant after
the loading. We measured the field noise from the
state-decomposition of a rf-dressed state (no Raman fields)
nominally on resonance and then feed-forward canceled the field
noise.  This reduced the $\sim 0.2\ \mu{\rm T}$ RMS magnetic field
noise at 60 Hz by about a factor of 20, and remaining RMS field
noise is $\sim 0.03\ \mu{\rm T}$ (including all frequency components
up to $\approx 5$~kHz). All of our measurements were performed by
locking to the 60 Hz line before loading into the dressed state.

\subsection*{3~~Momentum measurements of the dressed state }
The Raman-dressed state (no rf) is a superposition of spin and
momentum components; its canonical momentum $p_{\rm can}$ is the
momentum of the spin $m_F =0$ component. Experimentally, we fit the
$m_F=0$ density distribution after TOF to a Thomas-Fermi
profile\cite{Castin96} and identify $p_{\rm can}$ as the center of
the distribution. The mechanical momentum of the dressed state $mv$
was measured by a population-weighted average over all three spin
states including every pixel with discernable atoms in the image.
This takes into account the modification of the TOF density
distribution for all $m_F$ states due to interactions during TOF.
Although interactions can exchange momentum between spin states, the
total momentum is conserved.

\end{document}